\documentclass{mem}
\usepackage{natbib}\usepackage{txfonts}\usepackage{balance}
\usepackage{graphicx}
\usepackage[a4paper,breaklinks,dvipdfm]{hyperref}
\idline{1}{1}

\begin{document}
\def\kms{$\mathrm {km~s}^{-1}$}

\title{Testing the Connection Between Radio Mini-Halos and Core Gas Sloshing with MHD Simulations}

\author{
John ZuHone\inst{1}, 
Maxim Markevitch\inst{1,2},
\and Gianfranco Brunetti\inst{3}
}

\offprints{J. ZuHone}
 
\institute{
Harvard-Smithsonian Center for Astrophysics, Cambridge, MA, USA
\and
NASA/Goddard Space Flight Center, Greenbelt, MD, USA 
\and 
INAF – Istituto di Radioastronomia, Bologna, Italy \\
\email{jzuhone@cfa.harvard.edu}
}

\authorrunning{ZuHone, Markevitch, \& Brunetti}

\titlerunning{Radio Mini-Halos and Core Gas Sloshing}

\abstract{
Radio mini-halos are diffuse, steep-spectrum synchrotron sources
associated with a fraction of relaxed clusters of galaxies. Observations of some mini-halo sources indicate a correlation
between the radio emission and the X-ray signature of gas sloshing,
``cold fronts.'' Some authors have suggested turbulence associated
with the sloshing motions may reaccelerate relativistic electrons,
resulting in emission associated with the fronts. We present MHD simulations of core gas sloshing in a
galaxy cluster, where we measure the turbulence created by these
motions and employ passive tracer particles to act as relativistic
electrons that may be reaccelerated by such turbulence. Our preliminary
results support such a link between sloshing motions and particle reacceleration.
}
\maketitle{}

\section{Introduction}

Many galaxy clusters are sources of radio emission. One such class of
sources found in galaxy clusters are that of radio
mini-halos. Mini-halos are diffuse, steep-spectrum synchrotron sources found in the cores of so-called
``cool-core'' clusters. These sources typically are associated with a
central AGN and extend out to approximately the cooling radius of the
cluster gas \citep[for a review see][]{fer08}.

A number of mini-halos have emission that is correlated on the sky
with spiral-shaped ``cold fronts'' seen in the X-ray emission, believed to be the signature of
sloshing of the cluster's cool core gas \citep{MVF03,asc06}. \citet{maz08} discovered this correlation in two clusters, and
suggested that the correlation resulted from a population of
relativistic electrons that was reaccelerated by turbulence generated by the sloshing
motions. In order to determine whether or not the reacceleration efficiency
resulting from this turbulence is sufficient enough to reaccelerate
electrons and produce the corresponding radio emission, we have
performed MHD simulations of gas sloshing in a galaxy cluster with
tracer particles acting as the relativistic electrons.

\begin{figure*}[t!]
\resizebox{\hsize}{!}{\includegraphics[clip=true]{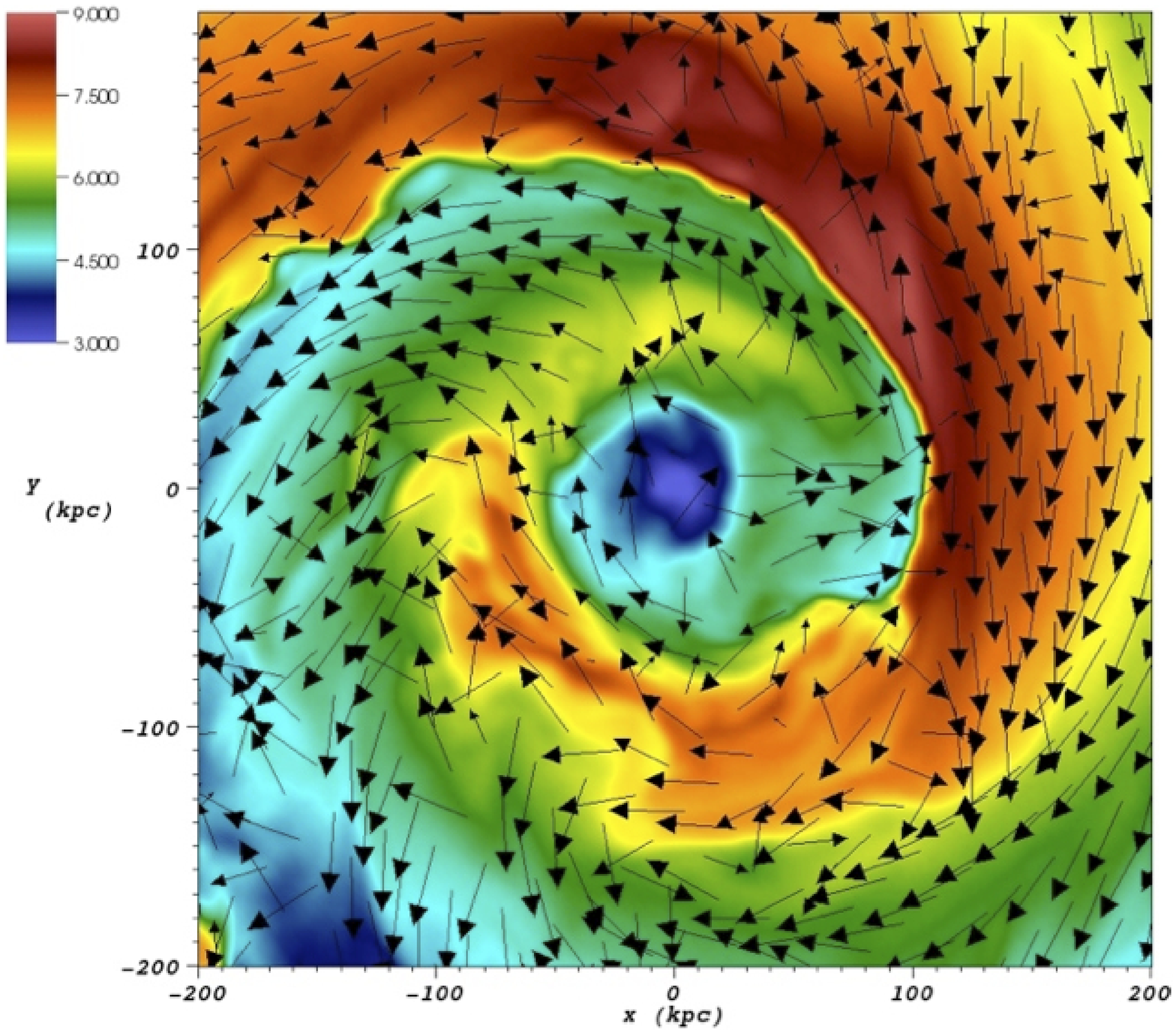}\includegraphics[clip=true]{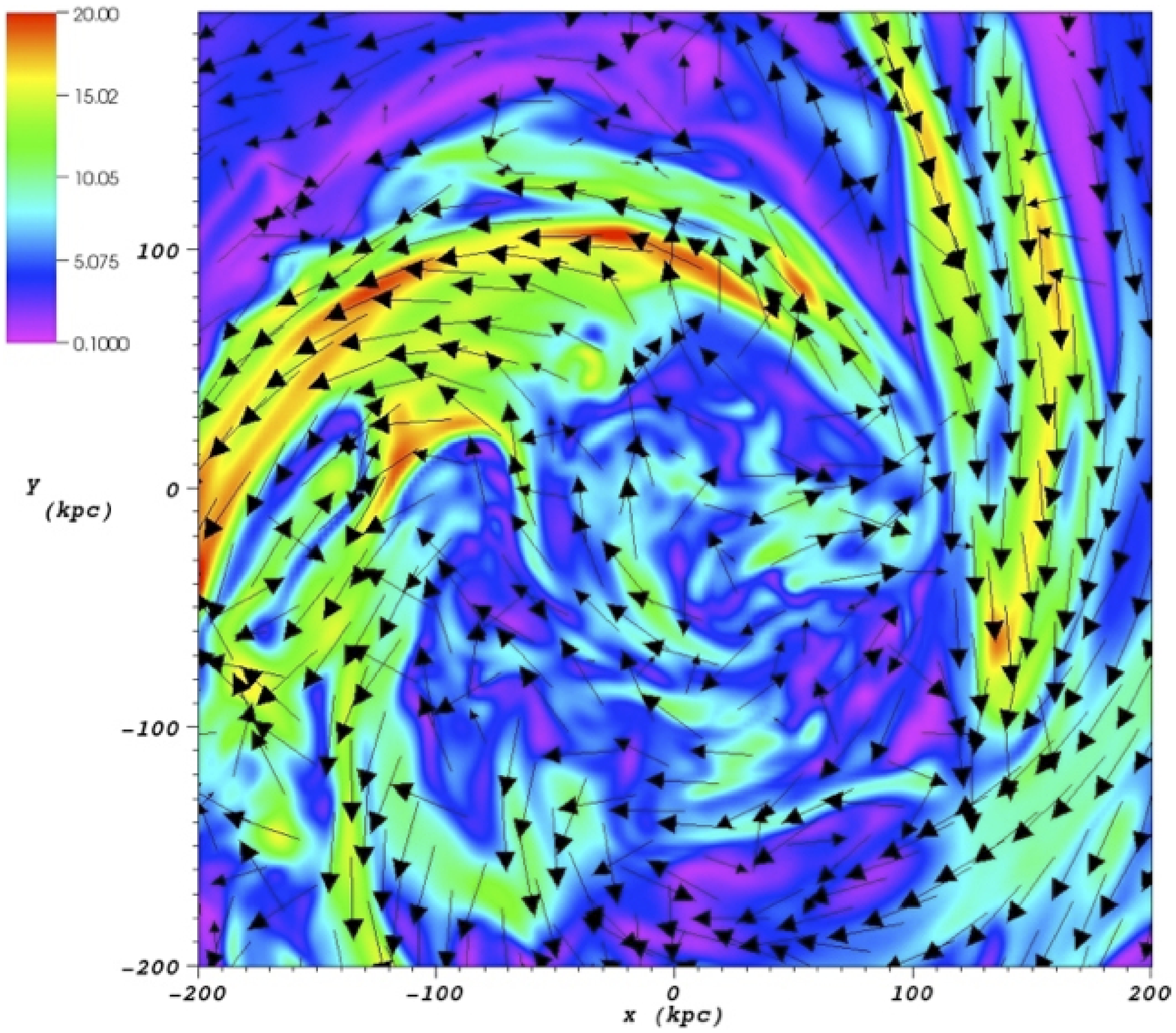}}
\caption{\footnotesize Slices through the center of the simulation
  domain of gas sloshing in the center of a galaxy cluster. Left:
  Temperature (keV), showing prominent spiral-shaped ``cold fronts''.
  Right: Magnetic field strength ($\mu$G), showing the field
  amplification near the front surfaces. Vectors show the magnetic
  field direction. Each panel is 400~kpc on a side. 
}
\label{fig:sloshing}
\end{figure*}

\section{Simulations}

Our simulations have been performed using FLASH 3, an adaptive mesh refinement hydrodynamics code with support for
simulations of magnetized fluids. Our simulations are set up in the manner of \citet{asc06} and
\citet{zuh10}. In this scenario, a large, relaxed cool-core cluster and a small, gasless
subcluster are set on a trajectory in which the subcluster will pass by the core. The gravitational force from
this subcluster acts on both the gas and dark matter cores of the main
cluster, but due to ram pressure the gas core becomes separated from
the center of the potential well and begins to ``slosh'' back and forth in the gravitational
potential, forming spiral-shaped cold fronts (see Figure
\ref{fig:sloshing}, left panel). The initial magnetic field in our simulations
is set up as a tangled field with an average field strength
proportional to the gas pressure ($\beta = p/p_B \approx$~100).

For the relativistic electrons, we employ a simple model, assuming
they are passive tracer particles advected along
with the fluid motions. Each particle is given an initial energy and carries with it the properties of the
local fluid. With this information, we derive the evolution of the
particle energy along its trajectories by taking into account the
relevant physical properties.

\begin{figure*}[t!]
\begin{center}
\includegraphics[width=0.5\textwidth]{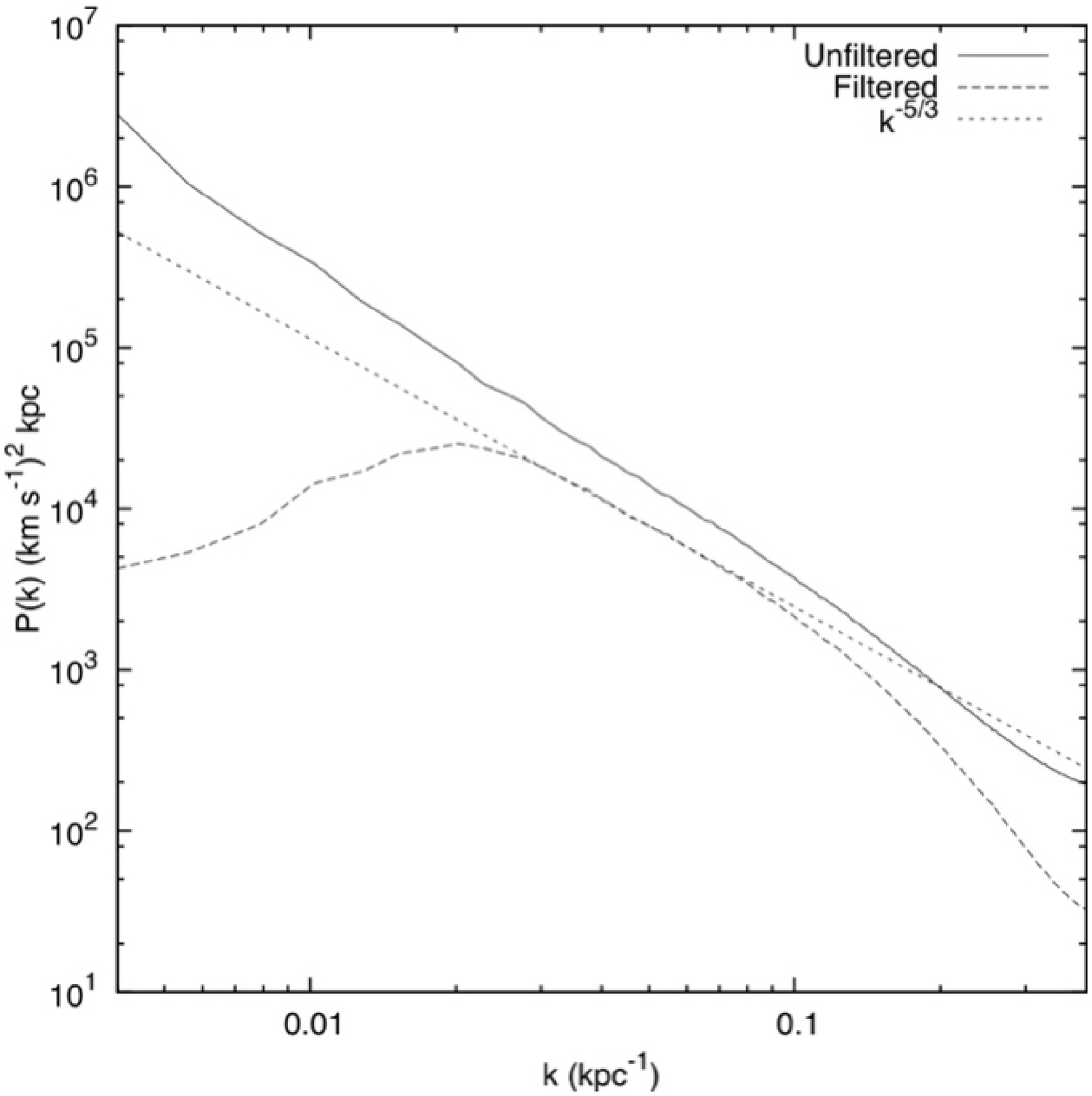}
\includegraphics[width=0.44\textwidth]{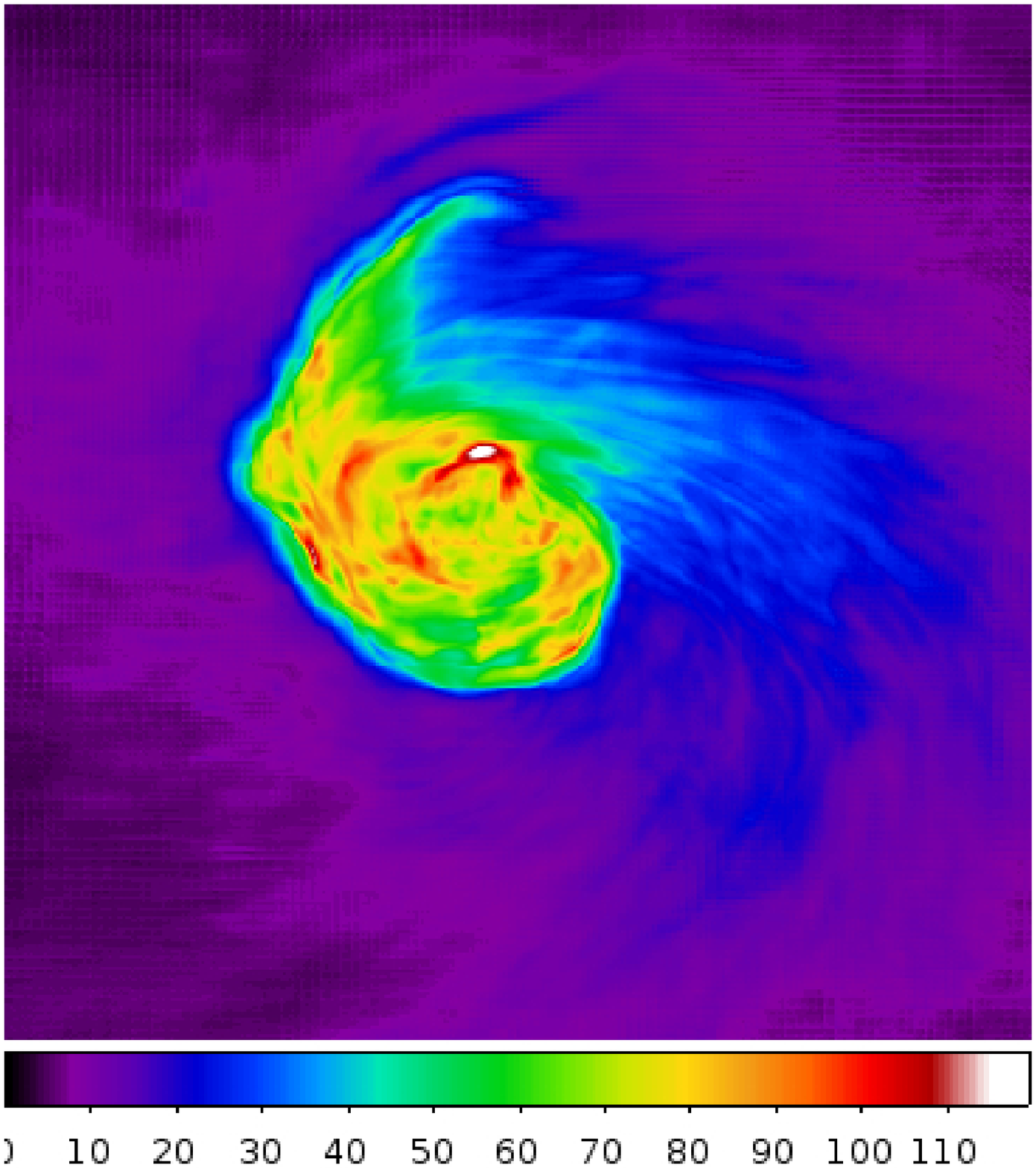}
\caption{\footnotesize The turbulent velocity structure of the
  sloshing gas. Left: The power spectrum of the velocity field. Solid
  line indicates the unfiltered power spectrum, dashed line the
  filtered power spectrum. The dotted line shows what would be
  expected for a Kolmogorov spectrum for comparison. Right:
  Mass-weighted, projected turbulent velocity of the gas in units of \kms. The most
  significant turbulence is contained within the envelope of the cold
  fronts. The panel is 400~kpc on a side. 
}
\label{fig:vels}
\end{center}
\end{figure*}

We assume that electrons are reaccelerated via transit-time damping
(TTD) of magnetosonic turbulent modes \citep{cas05,bru07,bru10}. We determine the turbulent velocity field in our simulation by a
process of filtering \citep{dol05,vaz06,vaz09}. The velocity field is
assumed to be a sum of a``bulk'' component and a ``turbulent''
component. For each position the local turbulent velocity is calculated by quadratically interpolating the mean
velocity from surrounding boxes of width $\sim$20~kpc and then
subtracting this mean value from the total velocity. The turbulent reacceleration coefficient for
the electrons then may be determined using the TTD formalism. Figure
\ref{fig:vels} shows the resulting turbulent velocity has a power spectrum
that is close to Kolmogorov (left panel) and is strongest in the
region of the sloshing motions (right panel). We take into account radiative (synchrotron and inverse Compton) and
Coulomb losses to calculate the evolution of the particle energy, with
the physical parameters determined from the properties of the
magnetized fluid. With the updated particle energies, we can then
compute the associated synchrotron emissivities that may be projected
along a line of sight to produce a map of radio emission.

\section{Results}

Two effects contribute to the association of the radio emission with
the sloshing cold fronts. Strong shear
flows are associated with the cold front surfaces, and these flows
stretch and amplify magnetic field lines parallel to these
surfaces \citep{kes10}. Our simulations show that the degree of amplification of the
magnetic energy along a cold front can be an order of magnitude or
more above the initial field energy (ZuHone et al. 2011, in
preparation; also see Figure \ref{fig:sloshing}, right panel). 

\begin{figure*}[t!]
\resizebox{\hsize}{!}{\includegraphics[clip=true]{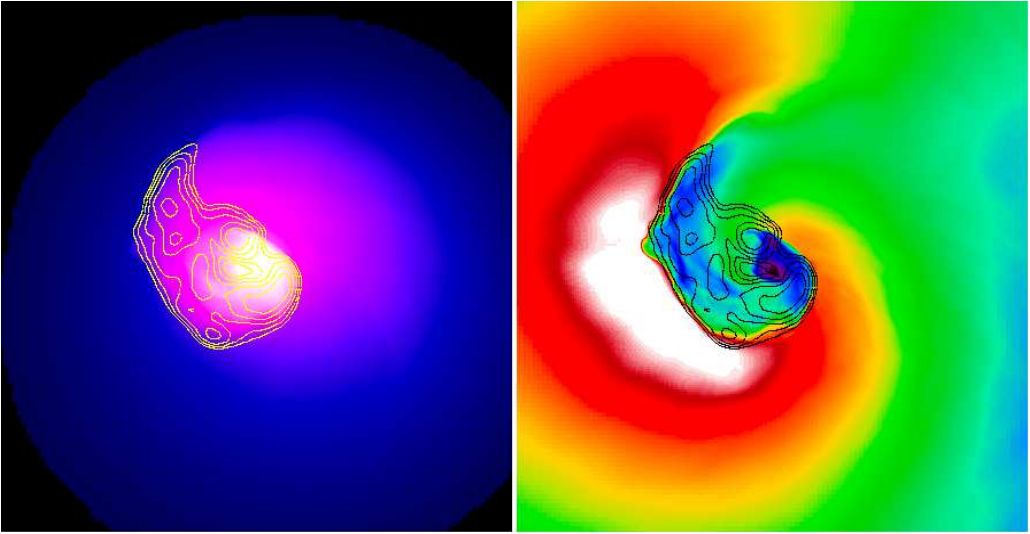}}
\caption{\footnotesize Mock X-ray images of a sloshing gas core with
  mock 300 MHz radio contours overlaid, projected along the z-axis of
  the simulation. Left: X-ray surface brightness in the 0.5-7.0
  keV band. Right: ``Spectroscopic-like'' projected temperature. Contours begin at 0.5 mJy/beam and are spaced
  by a factor of $\sqrt{2}$. Each frame is 400~kpc on a side. 
}
\label{fig:mini_halo}
\end{figure*}

Our most intriguing results come from the integration of electron
energies along the trajectories of our tracer particles. Beginning
with a spherical distribution of particles with radius $r$ = 200~kpc,
and with the number density proportional to the local gas density, we assign electron energies to
each tracer particle from an initial power-law distribution, set up
shortly after the sloshing period begins. After evolving the particle
energies along the trajectories of the tracer particles for
approximately 400~Myr, we find that most of the electrons have cooled
below the threshold for synchrotron emission, with the exception of
those associated with the envelope of the sloshing motions. Figure
\ref{fig:mini_halo} shows mock observations of X-ray surface
brightness and projected ``spectroscopic-like'' temperature with radio
contours overlaid. There is a clear correspondence with the region
associated with the cold fronts and the radio emission from the
mini-halo. This is also the region where the turbulent motions are
strongest (see Figure \ref{fig:vels}, right panel). If the electrons
are assumed to simply cool without reacceleration, we do not see such emission.

\section{Conclusions}

We have performed MHD simulations of gas sloshing in clusters of
galaxies, with the aim of determining whether or not the correlation
between radio mini-halo emission and sloshing cold fronts in some
clusters can be explained by the reacceleration of relativistic
electrons by turbulence associated with the sloshing motions. Our
initial results are very promising, indicating that the combined
effects of the amplified magnetic field and the turbulence associated
with the sloshing motions reaccelerates a population of relativistic
electrons within the envelope of the cold fronts which emit
synchrotron radiation from these regions. In our procedure we assume a
population of seed electrons diffused in the sloshing region. Although we do not model the injection process of these
particles, several mechanisms may provide an efficient source of fresh
electrons to reaccelerate in cluster cool-cores \citep[see discussion in][]{cas08}. Further work will
also detail differences in prediction with other models for mini-halos. 

\begin{acknowledgements}
J.A.Z. is supported under {\it Chandra} grant GO8-9128X. The software used in this work was in part developed by the DOE-supported ASC / Alliance Center for Astrophysical Thermonuclear Flashes at the University of Chicago.
\end{acknowledgements}

\bibliographystyle{aa}

\end{document}